\documentclass[12pt]{article}
\usepackage{amsmath}
\usepackage{amssymb}
\usepackage{fullpage}
\usepackage[dvips]{epsfig}

\def\##1{{\bf #1}}
\def\=#1{\underline{\underline{#1}}}

\def\eps{\epsilon}
\def\epso{\epsilon_0}
\def\muo{\mu_0}

\def\lambdao{\lambda_0}

\def\.{\mbox{ \tiny{$^\bullet$} }}

\def\les{\left[}
\def\ris{\right]}

\def\c#1{\cite{#1}}

\begin{document}

\begin{center}
{\large {\bf Reversal of Circular Bragg Phenomenon in Ferrocholesteric 
Materials with
Negative Real Permittivities and Permeabilities}}
\vskip 0.2cm

\noindent  { Akhlesh Lakhtakia} 
\vskip 0.2cm
\noindent {\em CATMAS~---~Computational \& Theoretical Materials Sciences Group\\
Department of Engineering Science \& Mechanics\\
212 Earth--Engineering Sciences Building\\
Pennsylvania State University, University Park, PA 16802--6812, USA}
\end{center}

\bigskip

{\sf A structurally right/left--handed ferrocholesteric slab with negative
real permittivities and permeabilities is theoretically shown
to display the Bragg phenomenon
on axial excitation
as if it were a structurally left/right--handed 
ferrocholesteric slab with positive
real permittivities and positive real permeabilities. In addition to the promise
of isotropic homo\-geneous materials with negative real permittivities and permeabilities
for un\-expected applications, the presented results underscore the similar potential
of  anisotropic non\-homo\-geneous materials with analogous characteristics.}

\bigskip

Dielectric and magnetic materials are ubiquitous. Their linear electromagnetic
response properties are characterized by permittivity and permeability
dyadics which depend on the frequency of excitation and comprise
complex--valued scalar components. The permittivity and the permeability
dyadics (i.e., second--order tensors)
of an isotropic material reduce to complex--valued scalars. The real
parts of these scalars can be negative or positive, but the latter possibility
is the normative one of the two, as the perusal of almost any undergraduate
electromagnetics textbook will show. However, the former possibility does
exist for natural materials such as metals, plasmas and ferrites \c{BH,DK}.
Most recently, composite materials
which effectively have both negative real permittivity and negative real
permeability in a certain frequency range have been fabricated and 
satisfactorily tested
\c{SPVNS,SSS}, notwithstanding the discounting of
 anisotropy, nonhomogeneity and dissipation in the sample mater\-ials
\c{Pen1}.

Technological bonanzas have been profferred,
provided homogeneous, isotropic and virtually non--dissipative materials
with negative permittivity and negative permeability can be economically 
manufactured \c{Pen1,Ves,Pen2}.
These potential benefits are based on the opposite directions of the
phase velocity and the velocity of energy transport in these materials.
Available results indicate that these materials would be realized
in the form of multilaminar slabs, each
lamina itself being anisotropic due to the imprinting
of various features thereon \c{SSS, Pen2}. Feature geometries
other than the only one in current use will also arise, sooner or later.
These open up the possibility of
an entirely new class of prospective materials: ferrocholesterics with negative permittivity and negative permeability. This Communication is devoted to these unidirectionally
nonhomogeneous materials.

About twenty years before the discovery of cholesteric liquid crystals by Reinitzer
in 1888 \c{Coll}, Reusch \c{Reu}
presented structurally similar materials made
from uniaxial dielectric laminas, which can possibly be
fabricated as unidirectional fibrous composites. The laminas are identical,
with the sole optic axis lying in the laminar plane. The laminas are stacked sequentially, the optic axis in any particular lamina offset by a small
angle in the laminar plane from the optic axis of the lamina
lying immediately below it. The successive optic axes rotate helicoidally,
and the optical response properties
of the entire structure resemble those of a cholesteric
liquid crystal at frequencies below a certain maximum \c{BS}. 
The optical response of cholesteric materials in the liquid crystalline form
has been intensively studied and technologically exploited \c{Bel,Jac}.

Just about
a century later, Brochard and de Gennes \c{BdeG} incorporated parallel
magnetic needles in the laminas, giving
rise to the so--called ferrocholesteric mater\-ials.
Let the thickness direction of a ferrocholesteric material
be parallel to the $x_3$ axis of a cartesian coordinate system
($x_1$, $x_2$, $x_3$). The simplest
effective constitutive equations of this material are as follows:
\begin{eqnarray}
&&\#D(\#x) = \epso \les \eps_a\,\=I + (\eps_b-\eps_a)\,\#c\#c\ris\.
\#E(\#x)\,,\\
&&\#B(\#x) = \muo \les \mu_a\,\=I + (\mu_b-\mu_a)\,\#c\#c\ris\.
\#H(\#x)\,.
\end{eqnarray}
Here, $\epso$ and $\muo$ are the permittivity and the permeability of
free space (i.e., vacuum); $\=I$ is the identity dyadic; the unit vector
\begin{equation}
\#c = \hat{\#x}_1 \,\cos(\pi x_3/\Omega) + h\,\hat{\#x}_2 \,\sin(\pi x_3/\Omega)
\end{equation}
involves the helicoidal pitch $2\Omega$; the parameter $h=+1$ for structural
right--handed\-ness, and $h=-1$ for structural left--handedness,
while the scalars $\eps_{a,b}$ and
$\mu_{a,b}$ are complex--valued functions of the angular frequency
$\omega$. 

As a ferrocholesteric material is periodically nonhomogeneous, it must
exhibit the circular Bragg phenomenon \c{LV}. Most significantly, if a 
circularly polarized plane wave
with angular frequency in the so--called Bragg regime were  normally incident
on a ferrocholesteric slab $0\leq x_3\leq L$ (which is thus
excited axially), and the ratio $L/\Omega$
were sufficiently high, it will be almost completely reflected if its handedness
matches the structural handedness of the slab. Virtually no reflection
would occur if the two handednesses do not coincide. The importance
of this polarization--discriminatory characteristic in optics cannot be
exaggerated, as it is exploited for a variety of polarization--sensitive
filters and laser mirrors \c{Jac}.

The reflectances and the transmittances of axially excited
ferrocholesteric slabs of infinite lateral extent were calculated by following
an established procedure \c{LW1}. These quantities were organized as
the $2\times 2$ matrixes
\begin{equation}
\les \begin{array}{cc} R_{RR} & R_{RL}\\ R_{LR} & R_{LL}
\end{array}\ris
\,,\qquad\qquad
\les \begin{array}{cc} T_{RR} & T_{RL}\\ T_{LR} & T_{LL}
\end{array}\ris\,.
\nonumber
\end{equation}
Here, the subscript RL indicates the intensity of either a reflected
or a transmitted {\em right} circularly polarized
plane wave in relation to the intensity of an
incident plane wave that is {\em left} circularly polarized; and so on.

In Figure 1 are shown the reflectances and the transmittances  
as functions of the free--space wavelength $\lambdao = 2\pi/\omega \sqrt{\epso\muo}$
for a  ferrocholesteric slab with the following properties:
$\eps_a = 3(1+0.001i)$, $\eps_b=3.3(1+0.001i)$, $\mu_a=1.2(1+0.002i)$,
$\mu_b=1.5 (1+0.002i)$, $h=1$, $\Omega = 14$~mm and $L=40\Omega$. 
Dispersion was ignored in this illustrative study, and an $\exp(-i\omega t)$ time--dependence was assumed. 
Differential reflection
of incident left/right circularly polarized plane waves is clearly in evidence in the
Bragg regime, which spans the wavelength range $\lambdao \in
\les 53.2,\,62.3\ris$~mm. As the chosen material is structurally right--handed,
$R_{RR}$ and $T_{LL}$
are enormously larger than the negligibly small $R_{LL}$ and $T_{RR}$
in the Bragg regime. The cross--polarized reflectances and transmittances~---~$R_{RL}$,
etc.,~---~are
also very small, and can be further reduced by the use of impedance--matching
layers.

The calculations for Figure 1 were repeated, except with
the following changes
in the input parameters: $\eps_a = -3(1-0.001i)$, $\eps_b=-3.3(1-0.001i)$, $\mu_a=-1.2(1-0.002i)$,
$\mu_b=-1.5 (1-0.002i)$. These values are in accord with the Lorentz oscillator
model \c{BH, SSS}, and therefore do not violate the principles
of causality and energy conservation. The calculated reflectance and transmittance spectrums are shown in Figure 2. The circular Bragg phenomenon
is preserved, but with a difference. Despite the fact that the chosen material
is {\em still\/} structurally right--handed, now $R_{LL}$ and $T_{RR}$
are enormously larger than the negligibly small $R_{RR}$ and $T_{LL}$.
The cross--polarized quantities remain unaffected.

In other words, a structurally right/left--handed ferrocholesteric 
slab with neg\-ative
real permittivities and negative real permeabilities will display the 
circular
Bragg phenomenon
on axial excitation
as if it were a structurally left/right--handed ferro\-cholesteric 
slab with positive
real permittivities and positive real permeabilities. Due to 
mathematical isomorphism,
this conclusion would also hold for ferrosmectic materials \c{LW2,SFJP}.

The reason underlying the foregoing symmetry can be understood by examining the
characteristics of the
axial propagation modes in ferrocholesteric and ferrosmectic materials;
detailed expressions are available elsewhere \c{LW2}. The propagation
modes are either left or right elliptically polarized, with their respective vibration ellipses
rotating along the $x_3$ axis in accordance with
the structural handedness
of the material \c{NLW}. Let dissipation be weak.
When the real parts ${\rm Re}\les \eps_{a,b}\ris >0$
and ${\rm Re}\les \mu_{a,b}\ris >0$, the direction of the phase velocity of a particular
mode is the same
as the (common) direction of energy transport and attenuation. However, when
${\rm Re}\les \eps_{a,b}\ris <0$
and ${\rm Re}\les \mu_{a,b}\ris <0$, not only does the phase velocity reverse in
direction, but the handedness of the vibration ellipse also reverses, while the direction
of energy flow and attenuation as well as 
the sense of rotation of the vibration ellipse remain unchanged. The reversal of
the modal handedness is thus responsible for the left/right switching between
the spectrums of Figures 1 and 2.

The Lorentz and the Drude models for oscillators are well--known \c{BH}.
These models yield negative as well as positive values for ${\rm 
Re}[\eps_{a,b},\,\mu_{a,b}]$ in different parts of the electromagnetic
spectrum, while the imaginary parts ${\rm Im}[\eps_{a,b},\,\mu_{a,b}]$ 
must be always positive. If the electric and the
magnetic resonances  are close to one another, the spectral regime of negative
permittivities and permeabilities may be accessible  simply by
increasing the frequency across all the resonances.

To conclude, many interesting phenomenons (such as anomalous refraction
and reversed Doppler shifts) and applications (such as distortion--free lenses)
have been forecasted for isotropic homogeneous materials with negative real
permittivity and permeability \c{SPVNS}--\c{Pen1}. The technology for fabricating these materials
can also be used for structurally nonhomogeneous materials with analogous
constitutive properties. Due to their anisotropic and nonhomogeneous constitution,
many unexpected phenomenons and applications are likely to emerge.

\newpage 
Figure Captions

\bigskip
Figure 1. Computed spectrums of the reflectances and transmittances
of a ferrocholesteric slab with ${\rm Re}\les \eps_{a,b},\, \mu_{a,b}\ris > 0$;
$\eps_a = 3(1+0.001i)$, $\eps_b=3.3(1+0.001i)$, $\mu_a=1.2(1+0.002i)$,
$\mu_b=1.5 (1+0.002i)$, $h=1$, $\Omega = 14$~mm and $L=40\Omega$.
Note that $R_{LR} = R_{RL}$ and $T_{LR}=T_{RL}$, correct to graphical
accuracy.

\bigskip

Figure 2. Same as Figure 1, but for 
$\eps_a = -3(1-0.001i)$, $\eps_b=-3.3(1-0.001i)$, $\mu_a=-1.2(1-0.002i)$, and
$\mu_b=-1.5 (1-0.002i)$.


\begin{thebibliography}{99}

\bibitem{BH}
C.F. Bohren, D.R. Huffman,
{\em Absorption and Scattering of Light by Small Particles\/},
Wiley, New York {\bf 1998}.

\bibitem{DK}
R.F. Soohoo,
{\em Theory and Application of Ferrites\/},
Prentice--Hall, Englewood Cliffs, NJ, USA {\bf 1960}.


\bibitem{SPVNS}
D.R. Smith, W.J. Padilla, D.C. Vier, S.C. Nemat--Nasser, S. Schultz,
{\em Phys. Rev. Lett.\/} {\bf 2000}, {\em 84},  4184.

\bibitem{SSS}
R.A. Shelby, D.R. Smith, S. Schultz,
{\em Science\/} {\bf 2001}, {\em 292}, 77.

\bibitem{Pen1}
J.B. Pendry,
{\em Phys. Rev. Lett.\/} {\bf 2000}, {\em 85}, 3966.
[See also
correspondence on this paper: G.W. 't Hooft, {\em Phys. Rev. Lett.\/} {\bf 
2001}, {\em 87}, 249701;
J. Pendry, {\em Phys. Rev. Lett.\/} {\bf 
2001}, {\em 87}, 249702;
J.M. Williams, {\em Phys. Rev. Lett.\/} {\bf 
2001}, {\em 87}, 249703;
J. Pendry, {\em Phys. Rev. Lett.\/} {\bf 
2001}, {\em 87}, 249704.]

\bibitem{Ves}
V.S. Veselago,
{\em Usp. Fiz. Nauk.\/} {\bf 1967}, {\em 92}, 517;
{\em Sov. Phys. Usp.\/} {\bf 1968}, {\em 10}, 509.


\bibitem{Pen2}
J.B. Pendry,
{\em Physics World\/} {\bf 2001}, {\em 14\/}(5), 47; September issue.

\bibitem{Coll}
P.J. Collings,
{\em Liquid Crystals, Nature's Delicate Phase of Matter\/},
Princeton University Press, Princeton, NJ,  USA {\bf 1990}.

\bibitem{Reu}
E. Reusch,
{\em Ann. Phys. Chem.\/} {\bf 1869}, {\em 138}, 628.

\bibitem{BS}
D.W. Berreman, T.J. Scheffer,
{\em Mol. Cryst. Liq. Cryst.\/} {\bf 1970}, {\em 11}, 395.

\bibitem{Bel}
V.A. Belyakov,
{\em Diffraction Optics of Complex--Structured Periodic Media\/},
Springer, New York, NY, USA {\bf 1992}.

\bibitem{Jac}
S.D. Jacobs (Ed.),
{\em Selected Papers on Liquid Crystals for Optics\/},
SPIE, Bellingham, WA,  USA {\bf 1992}.


\bibitem{BdeG}
F. Brochard, P.G. de Gennes,
{\em J. Phys. (Paris)\/} {\bf 1970}, {\em 31}, 691.

\bibitem{LV}
A. Lakhtakia, V.C. Venugopal,
{\em Arch. Elektron. \"Uber.\/} {\bf 1999}, {\bf 53}, 287.


\bibitem{LW1}
A. Lakhtakia, W.S. Weiglhofer,
{\em Proc. R. Soc. Lond. A\/} {\bf 1995}, {\em 448}, 419; {\bf 1998}, {\em 454},
3275.

\bibitem{LW2}
A. Lakhtakia, W.S. Weiglhofer,
{\em Opt. Commun.\/} {\bf 1994}, {\em 111}, 199; {\bf  1995}, {\em 113}, 570.

\bibitem{SFJP}
D. Spolianski, J. Ferre, J.P. Jamet, V. Ponsinet,
{\em J. Mag. Mag. Mater.\/} {\bf 1999}, {\em 201}, 200.

\bibitem{NLW}
S.F. Nagle, A. Lakhtakia, W. Thompson, Jr.,
{\em J. Acoust. Soc. Am.\/} {\bf 1995}, {\em 97}, 42.

\end{thebibliography}
\end{document}